\documentclass[twocolumn,amsmath,amssymb,aps,preprintnumbers,prl]{revtex4-1}
\usepackage{times}
\usepackage{amsmath}
\usepackage{amsfonts}
\usepackage{graphicx}
\usepackage{dcolumn}
\usepackage{bm}
\usepackage{color}

\def\beq{\begin{equation}}
\def\eeq{\end{equation}}
\def\bea{\begin{eqnarray}}
\def\eea{\end{eqnarray}}
\def\beqn{\begin{eqnarray}} \def\eeqn{\end{eqnarray}}
\def\beeq{\begin{eqnarray}}
\def\eeeq{\end{eqnarray}}

\def\nn{\nonumber}
\def\Eq#1{Eq.~(\ref{#1})}

\def\td#1{\tilde{\delta}\left(#1\right)}

\def\qon#1{q_{#1,0}^{(+)}}

\def\im#1{{\rm Im}(#1)}

\def\res#1{{\rm Res}\left(#1\right)}

\def\qb{\mathbf{q}}

\def\im{{\rm Im}}

\def\ii{\imath 0}

\begin{document}

\preprint{IFIC/20-02}

\title{Open loop amplitudes and causality to all orders and powers from the loop-tree duality}

\author{J. Jes\'us Aguilera-Verdugo~$^{(a)}$} 
\author{F\'elix Driencourt-Mangin~$^{(a)}$} 
\author{Roger J. Hern\'andez-Pinto~$^{(b)}$} 
\author{Judith Plenter~$^{(a)}$} 
\author{Selomit Ram\'{\i}rez-Uribe~$^{(a,b,c)}$} 
\author{Andr\'es E. Renter\'{\i}a-Olivo~$^{(a)}$} 
\author{Germ\'an Rodrigo~$^{(a)}$} 
\author{Germ\'an F. R. Sborlini~$^{(a)}$} 
\author{William J. Torres Bobadilla~$^{(a)}$} 
\author{Szymon Tracz~$^{(a)}$} 
\affiliation{ ${}^{a}$ Instituto de F\'{\i}sica Corpuscular, Universitat de Val\`{e}ncia -- Consejo Superior de Investigaciones Cient\'{\i}ficas, 
Parc Cient\'{\i}fic, E-46980 Paterna, Valencia, Spain. \\
${}^{b}$ Facultad de Ciencias F\'{\i}sico-Matem\'aticas,
Universidad Aut\'onoma de Sinaloa, Ciudad Universitaria, CP 80000 Culiac\'an, Mexico. \\
${}^{c}$ Facultad de Ciencias de la Tierra y el Espacio,
Universidad Aut\'onoma de Sinaloa, Ciudad Universitaria, CP 80000 Culiac\'an, Mexico. }

\date{May 1, 2020}

\begin{abstract}
Multiloop scattering amplitudes describing the quantum fluctuations at high-energy scattering processes 
are the main bottleneck in perturbative quantum field theory. 
The loop-tree duality is a novel method aimed at overcoming this bottleneck by opening the loop amplitudes into trees and 
combining them at integrand level with the real-emission matrix elements. 
In this Letter, we generalize the loop-tree duality to all orders in the perturbative expansion
by using the complex Lorentz-covariant prescription of the original one-loop formulation. 
We introduce a series of mutiloop topologies with arbitrary internal configurations
and derive very compact and factorizable expressions of their open-to-trees representation in the loop-tree 
duality formalism. Furthermore, these expressions are entirely independent at integrand level of the initial assignments
of momentum flows in the Feynman representation and remarkably free of noncausal singularities.
These properties, that we conjecture to hold to other topologies at all orders, provide integrand representations 
of scattering amplitudes that exhibit manifest causal singular structures and better numerical stability than in other representations.   
\end{abstract}

\maketitle


\section{Introduction}

Precision modeling of fundamental interactions relies mostly on perturbative quantum field theory. 
Quantum fluctuations in perturbative quantum field theory are encoded by Feynman diagrams with closed loop circuits. These loop diagrams are 
the main bottleneck to achieve higher perturbative orders and therefore more precise theoretical predictions
for high-energy colliders~\cite{hllhc,Abada:2019lih}. 
Whereas loop integrals are defined in the Minkowski space of the loop four-momenta, the loop-tree duality (LTD)~\cite{Catani:2008xa,Bierenbaum:2010cy,Bierenbaum:2012th,Buchta:2014dfa,Buchta:2015wna,Buchta:2015xda,Hernandez-Pinto:2015ysa,Sborlini:2016gbr,Sborlini:2016hat,Driencourt-Mangin:2017gop,Driencourt-Mangin:2019aix,Driencourt-Mangin:2019sfl,Aguilera-Verdugo:2019kbz,Driencourt-Mangin:2019yhu,Plenter:2019jyj,Tomboulis:2017rvd,Runkel:2019yrs,Runkel:2019zbm,Capatti:2019ypt,Capatti:2019edf}
exploits the Cauchy residue theorem to reduce the dimensions of the integration domain by one unit in each loop. 
In the most general version of LTD the loop momentum component that is integrated out is arbitrary~\cite{Catani:2008xa,Bierenbaum:2010cy}. 
In numerical implementations~\cite{Buchta:2015wna,Buchta:2015xda,Sborlini:2016gbr,Sborlini:2016hat,Driencourt-Mangin:2019aix,Driencourt-Mangin:2019sfl,Driencourt-Mangin:2019yhu,Capatti:2019ypt,Capatti:2019edf}
and asymptotic expansions~\cite{Driencourt-Mangin:2017gop,Plenter:2019jyj},
it is convenient to select the energy component because the remaining integration domain, the loop three-momenta, is Euclidean.

LTD opens any loop diagram to a forest (a sum) of nondisjoint trees by introducing as many on-shell conditions on the 
internal loop propagators as the number of loops, and is realized by modifying the usual infinitesimal complex
prescription of the Feynman propagators. The new propagators with modified prescription are called dual propagators. 
LTD at higher orders proceeds iteratively, or in the words of Feynman~\cite{Feynman:1963ax,Feynman:1972mt}, 
by {\it opening the loops in succession}.
While the position of the poles of Feynman propagators in the complex plane is well defined, i.e., the positive (negative) energy modes 
feature a negative (positive) imaginary component due to the momentum independent $+\ii$ imaginary prescription, 
the dual prescription of dual propagators is momentum dependent. Therefore, after applying LTD to the first loop, 
the position of the poles in the complex plane of the subsequent loop momenta is modified. 
The solution found in Refs.~\cite{Bierenbaum:2010cy,Bierenbaum:2012th} was to reshuffle the imaginary components 
of the dual propagators by using a general identity that relates dual with Feynman propagators in such a way 
that propagators entering the second and successive applications of LTD are Feynman propagators only. 
This procedure requires to reverse the momentum flow of a few subsets of propagators 
in order to keep a coherent momentum flow in each LTD round.

\begin{figure}[th]
\begin{center}
\includegraphics[width=0.45\textwidth]{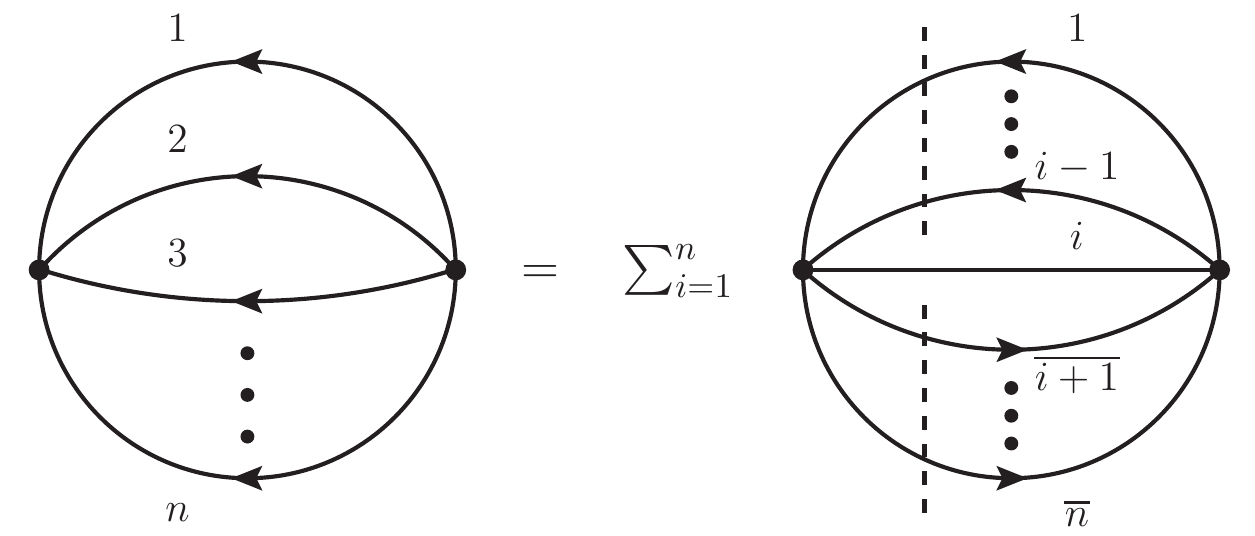}
\caption{Maximal loop topology (left) and the corresponding open dual representation (right).  
An arbitrary number of external legs is attached to each loop line.
All the propagators in the set $i$ on the rhs are off shell, while 
the dashed line represents the on-shell cut over the other $n-1$ sets: one on-shell propagator in each 
set and an implicit sum over all possible on-shell configurations.
Bars indicate a reversal of the momentum flow.
\label{fig:maxtop}}
\end{center}
\end{figure}

Recent papers have proposed alternative dual representations~\cite{Runkel:2019yrs,Runkel:2019zbm,Capatti:2019ypt,Capatti:2019edf}.
In Refs.~\cite{Runkel:2019yrs,Runkel:2019zbm}, an average of all the possible momentum flows is proposed, which requires 
a detailed calculation of symmetry factors. We show in this Letter that this average is unnecessary.
In Refs.~\cite{Capatti:2019ypt,Capatti:2019edf}, the Cauchy residue theorem is applied iteratively 
by keeping track of the actual position of the poles in the complex plane. The procedure requires to 
close the Cauchy contours at infinity from either below or above the real axis, in order to cancel the 
dependence on the position of the poles.
In this Letter, we follow a new strategy to generalize LTD to all orders, and with the original Lorentz-covariant prescription~\cite{Catani:2008xa,Bierenbaum:2010cy}.
As in Refs.~\cite{Bierenbaum:2010cy,Bierenbaum:2012th,Driencourt-Mangin:2019aix,Driencourt-Mangin:2019sfl}, 
we reverse sets of internal momenta whenever
it is necessary to keep a coherent momentum flow, and we close the Cauchy contours always in the lower complex half-plane. 
Causality~\cite{Buchta:2014dfa,Aguilera-Verdugo:2019kbz,Cutkosky:1960sp,Mandelstam:1960zz,Landau:1959fi,Cutkosky:1961,Coleman:1965xm,Kershaw:1971rc}
is also used as a powerful guide to select which kind of dual contributions 
are endorsed, and then construct suitable {\it Ans\"atze} that are proven by induction.  
This procedure allows us to obtain explicit and very compact analytic expressions of the 
LTD representation for a series of loop topologies to all orders and arbitrary internal configurations. 

\section{Loop-tree duality to all orders and powers}

The internal propagators of any multiloop integral or scattering amplitude can be classified into different sets or loop lines, 
each set collecting all the propagators that depend on the same single loop momentum or a linear combination of them. 
To simplify the notation,  $s$ labels the set of all the internal propagators $i_s \in s$ carrying momenta of the form $q_{i_s}=\ell_s+k_{i_s}$,
where $\ell_s$ is the loop momentum identifying this set, and where $k_{i_s}$ is a linear combination of external momenta $\left\{p_j\right\}_N$. 
Note that $\ell_s$ may be a linear combination of loop momenta, so long as it is the same fixed combination in all the elements in the set $s$.
The usual Feynman propagator of one single internal particle is 
\beq
G_F(q_{i_s}) = \frac{1}{q_{i_s,0}^2 -  \left(\qon{i_s}\right)^2}~,
\eeq
where 
\beq
q_{i_s,0}^{(+)} = \sqrt{\qb_{i_s}^2+m_{i_s}^2-\ii}~, 
\eeq
with $q_{i_s,0}$ and $\qb_{i_s}$ the time and spatial components of the momentum $q_{i_s}$, 
respectively, $m_{i_s}$ its mass, and $\ii$ the usual Feynman's infinitesimal imaginary prescription. 
We extend this definition to encode in a compact way the product of the Feynman propagators of one set or 
the union of several sets:
\beq
G_F(1,\ldots, n) = \prod_{i\in 1\cup \cdots \cup n} \left( G_F(q_i) \right)^{a_i}~.
\eeq
Here, we contemplate the general case where the Feynman propagators are raised to arbitrary 
powers. Still, the powers $a_i$ will appear only implicitly in the following. 
A typical $L$-loop scattering amplitude is expressed as 
\beq
{\cal A}^{(L)}_N(1,\dots,n) = \int_{\ell_1,\ldots, \ell_L} {\cal N}(\{\ell_i \}_L, \{p_j\}_N) \, G_F(1, \ldots, n)
\label{eq:typical}
\eeq
in the Feynman representation, i.e. as an integral in the Minkowski space of the $L$-loop momenta 
over the product of Feynman propagators and the numerator ${\cal N}(\{\ell_i \}_L, \{p_j\}_N)$, which 
is given by the Feynman rules of the theory. The integration measure reads $\int_{\ell_i} = -\imath \, \mu^{4-d} \int d^d\ell_i/(2\pi)^d$
in dimensional regularization~\cite{Bollini:1972ui,tHooft:1972tcz}, with $d$ the number of space-time dimensions.

Beyond one loop, any loop subtopology involves at least two loop lines that depend on the same loop momentum. 
We define the dual function that accounts for the sum of residues in the complex plane 
of the common loop momentum as
\beq
G_D(s; t) = -2\pi \imath \sum_{i_s \in s} \res{G_F(s,t), \im(\eta \cdot q_{i_s}) < 0}~,
\label{eq:GDst}
\eeq
where $G_F(s,t)$ represents the product of the Feynman propagators that belong 
to the two sets $s$ and $t$. Each of the Feynman propagators can be raised to an arbitrary power. 
Notice that in \Eq{eq:GDst} only the propagators that belong to the set $s$ are set consecutively 
on shell. The Cauchy contour is always closed from below the real axis,  $\im(\eta \cdot q_{i_s}) < 0$. 
The vector $\eta$ is futurelike and was introduced in the original formulation of 
LTD~\cite{Catani:2008xa} to regularize the dual propagators in a Lorentz-covariant form.
For single power propagators, $s=t$ and $\eta=(1,{\bf 0})$,
\Eq{eq:GDst} provides the customary dual function at one loop with the energy component integrated out
\beq
G_D(s) = - \sum_{i_s\in s} \td{q_{i_s}} \prod_{\stackrel{j_s\ne i_s}{j_s\in s }}\frac{1}{(\qon{i_s}+ k_{j_s i_s,0})^2 - (\qon{j_s})^2}~,
\eeq
with $k_{j_s i_s} = q_{j_s}-q_{i_s}$, and $\td{q_{i_s}}= 2\pi \imath \, \theta(q_{i_s,0}) \delta(q_{i_s}^2-m_{i_s}^2)$
selecting the on-shell positive energy mode, $q_{i_s,0}>0$. If some of the Feynman propagators are raised to 
multiple powers, then \Eq{eq:GDst} leads to heavier expressions~\cite{Bierenbaum:2012th} but the location of the 
poles in the complex plane is the same as in the single power case.  

Then, we construct the dual function of nested residues involving several sets of momenta
\bea
&& G_D(1, \ldots, r; n) =  -2\pi \imath \nn \\ &&\times \sum_{i_r \in r} \res{G_D(1, \dots, r-1; r, n), \im(\eta \cdot q_{i_r})<0}~. 
\label{eq:GDn}
\eea
In the rhs of Eqs.~(\ref{eq:GDst}) and (\ref{eq:GDn}), we can introduce numerators or replace the Feynman propagators
by the integrand of \Eq{eq:typical} to define the corresponding unintegrated open dual amplitudes ${\cal A}^{(L)}_D(1,\ldots,r; n)$.
An example of dual amplitudes at two loops was presented in Ref.~\cite{Driencourt-Mangin:2019aix}.

In the next sections, we will derive the LTD representation of the multiloop scattering amplitude 
in \Eq{eq:typical} and we will present explicit expressions for several benchmark topologies to all orders.
The notation introduced above allows us to express the LTD representations in a very compact way, since
it only requires to label and specify the overall structure of the loop sets, regardless of their internal and specific configuration.

\section{Maximal loop topology}

The maximal loop topology (MLT), see Fig.~\ref{fig:maxtop},  is defined by $L$-loop topologies 
with $n=L+1$ sets of propagators, where the momenta of the propagators belonging to the first $L$ sets
depend on one single loop momentum, $q_{i_s} = \ell_s + k_{i_s}$ with $s\in\{1,\ldots,L\}$, 
and the momenta of the extra set, denoted by $n$, are a linear 
combination of all the loop momenta, $q_{i_n} = -\sum_{s=1}^{L} \ell_s + k_{i_n}$.
The minus sign in front of the sum is imposed by momentum conservation. 
The momenta $k_{i_s}$ and $k_{i_n}$ are linear combinations of external momenta. 
At two loops ($n=3$),  this is the only possible topology, and therefore 
sufficient to describe any two-loop scattering amplitude.

The LTD representation of the multiloop MLT amplitude, starting at two loops, is
extremely simple and symmetric
\bea
&& {\cal A}^{(L)}_{\rm MLT}(1,\dots,n) \nn \\ &&= \int_{\ell_1,\ldots, \ell_L} 
 \sum_{i=1}^{n} {\cal A}^{(L)}_D(1, \ldots, i-1, \overline{i+1}, \ldots, \overline n ;  i)~,
\label{eq:MLT}
\eea
with ${\cal A}^{(L)}_D(\overline 2, \ldots, \overline n; 1)$ and ${\cal A}^{(L)}_D(1, \ldots, n-1; n)$ 
as the first and the last elements of the sum, respectively.
The bar in $\overline s$ indicates that the momentum flow of the set $s$ is reversed ($q_{i_s}\to -q_{i_s}$), which is equivalent
to selecting the on-shell modes with negative energy of the original momentum flow. 
The compact expression in \Eq{eq:MLT} was obtained by first evaluating the nested residues, \Eq{eq:GDn}, of several representative 
multiloop integrals. The derived expressions were then used to formulate an {\it Ansatz} to all orders that
was proven by induction. It is noteworthy that there is no dependence in this expression on the position of the poles in the complex plane. 

In each term of the sum in the integrand of \Eq{eq:MLT} there is one set $i$ with all its propagators off shell, 
and there is one on-shell propagator in each of the other $n-1$ sets. 
This is the necessary condition to open the multiloop amplitude into nondisjoint trees.
Note also that there is an implicit sum over all possible on-shell configurations of the $n-1$ sets. 
The LTD representation in \Eq{eq:MLT} is displayed graphically in Fig.~\ref{fig:maxtop},
and represents the basic building block entering other topologies.

The causal behavior of \Eq{eq:MLT} is also clear and manifest. 
The dual representation in \Eq{eq:MLT} becomes singular when one or more off-shell propagators 
eventually become on shell and generate a disjoint tree dual subamplitude. 
If these propagators belong to a set where there is already one on-shell propagator 
then the discussion reduces to the one-loop case~\cite{Buchta:2014dfa}. We do not comment further on this case. The
interesting case occurs when the propagator becoming singular belongs to
the set with all the propagators off shell~\cite{Aguilera-Verdugo:2019kbz}. 
For example, the first element of the sum in \Eq{eq:MLT} features all the propagators in the set $1$ off shell. One of 
those propagators might become on shell, and there are two potential singular solutions, one with positive energy and another
with negative energy, depending on the magnitude and direction of the external momenta~\cite{Buchta:2014dfa,Aguilera-Verdugo:2019kbz}. 
The solution with negative energy represents a singular configuration where there is 
at least one on-shell propagator in each set.  Therefore, the amplitude splits into two disjoint trees, 
with all the momenta over the causal on-shell cut pointing to the same direction. Abusing notation:
\beq
{\cal A}_D^{(L)} (\overline 2, \ldots, \overline n; 1) \stackrel{1~\rm{on-shell}}{\to}{\cal A}_D^{(L)} (\overline 1, \overline 2, \ldots, \overline n)~.
\eeq
The on-shell singular solution with positive energy, however, is locally entangled with the next term in \Eq{eq:MLT} such that 
the full LTD representation remains nonsingular in this configuration:
\bea
&&{\cal A}_D^{(L)} (\overline 2, \overline 3, \ldots, \overline n; 1) + {\cal A}_D^{(L)} (1, \overline 3, \ldots, \overline n; 2)    \\
&& \stackrel{(1,2)~\rm{on-shell}}{\to}  {\cal A}_D^{(L)} (1, \overline 2, \overline 3, \ldots, \overline n) - {\cal A}_D^{(L)} (1, \overline 2, \overline 3, \ldots, \overline n)~.
\nn \eea
These local cancellations also occur with multiple power propagators.
They are the known dual cancellations of unphysical or noncausal 
singularities~\cite{Buchta:2014dfa,Driencourt-Mangin:2019aix,Driencourt-Mangin:2019sfl,Aguilera-Verdugo:2019kbz}
and their cancellation is essential to support that the remaining causal and anomalous 
thresholds as well as infrared singularities are restricted to a compact region of the loop three-momenta.
Causality determines that the only surviving singularities fall on ellipsoid surfaces in the loop three-momenta 
space~\cite{Buchta:2015wna,Buchta:2015xda,Capatti:2019edf}, that collapse to finite segments for massless particles 
leading to infrared singularities.  These causal singularities are bounded by the magnitude of the external momenta,
thus enabling the simultaneous generation with the tree contributions describing emission of extra radiation
through suitable momentum mappings, as defined in 
four-dimensional unsubstraction (FDU)~\cite{Hernandez-Pinto:2015ysa,Sborlini:2016gbr,Sborlini:2016hat}.
Another potential causal singularity occurs from the last term in \Eq{eq:MLT}
when all the on-shell momenta are aligned in the opposite direction over the causal on-shell cut,
${\cal A}_D^{(L)} (1, \ldots, n-1; n) \stackrel{n~\rm{on-shell}}{\to} {\cal A}_D^{(L)} (1, \ldots, n)$.

It is also interesting to note the remarkable structure that the LTD representation exhibits when  
expressed in terms of dual propagators. For example, the scalar MLT integral with 
only one single propagator in each set, e.g. the sunrise diagram at two loops, reduces to the extremely compact expression
\beq
{\cal  A}^{(L)}_{\rm MLT}(1,\ldots,n) = - \int_{\vec \ell_1, \ldots, \vec \ell_L} \frac{1}{2\qon{n}}
\left( \frac{1}{\lambda^-_{1,n}} + \frac{1}{\lambda^+_{1,n}} \right)~,
\label{eq:sunrise}
\eeq
where $\lambda^\pm_{1,n} =  \sum_{i=1}^{n} \qon{i} \pm k_{0,n}$, with $k_n = \sum_{i=1}^{n} q_i $, 
and $\int_{\vec \ell_s} = - \mu^{4-d} \, \int d^{d-1} \ell_s / (2\pi)^{d-1} / (2 \qon{s})$.
The most notable property of this expression is that it is explicitly free of unphysical singularities, and the causal singularities 
occur, as expected, when either $\lambda^+_{1,n}$ or $\lambda^-_{1,n}$ vanishes, depending on the sign of the energy component of $k_n$,
in the loop three-momenta region where the on-shell energies are bounded, $\qon{i}< |k_{0,n}|$.
This property also holds for powered propagators, nonscalar integrals, and more than one propagator in each set.
Furthermore, \Eq{eq:sunrise} is independent of the initial momentum flows in the Feynman representation.

\begin{figure}[th]
\begin{center}
\includegraphics[width=0.45\textwidth]{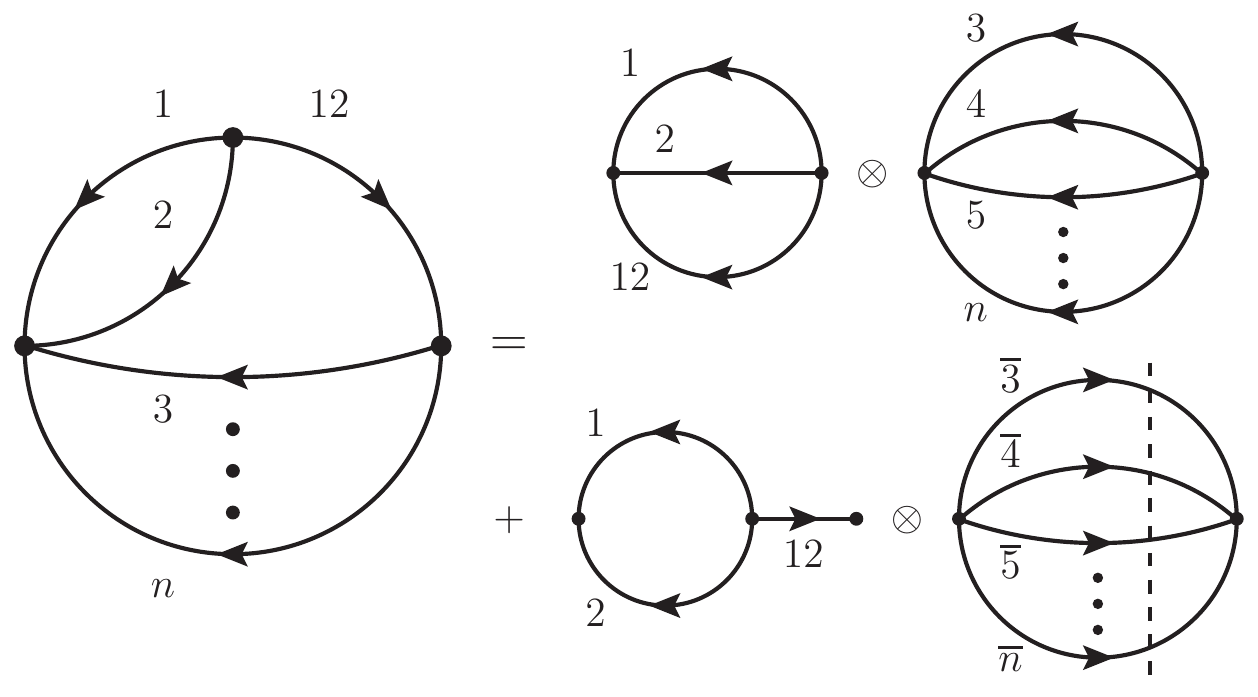} \qquad
\caption{Next-to-maximal loop topology (left) and its convoluted dual representation (right). 
Each MLT subtopology opens according to \Eq{eq:MLT}. Only the on-shell cut of 
the last MLT-like subtopology with reversed momentum flow is shown.
\label{fig:nmaxtop}}
\end{center}
\end{figure}

\section{Next-to-Maximal loop topology}

The next multiloop topology in complexity, see Fig.~\ref{fig:nmaxtop}, contains one extra set 
of momenta, denoted by $12$, that depends on the sum of two loop momenta, $q_{i_{12}} = - \ell_1 - \ell_2 + k_{i_{12}}$.
We call it next-to-maximal loop topology (NMLT). This topology appears for the first time at three loops, i.e. $n+1$ sets with $L=n-1$ and $n\ge 4$,
and its LTD representation is given by the compact and factorized expression
\bea
&& {\cal A}^{(L)}_{\rm NMLT}(1,\dots,n,12) 
= {\cal A}^{(2)}_{\rm MLT}(1, 2, 12) \otimes {\cal A}^{(L-2)}_{\rm MLT}(3, \ldots, n) \nn \\ 
&&\qquad + {\cal A}^{(1)}_{\rm MLT}(1,2)\otimes {\cal A}^{(0)}(12)\otimes  {\cal A}^{(L-1)}_{\rm MLT}(\overline 3,\ldots, \overline n)~.
\label{eq:NMLT}
\eea
The first term on the rhs of \Eq{eq:NMLT} represents a convolution of the two-loop MLT subtopology 
involving the sets $(1,2,12)$ with the rest of the amplitude, which is also MLT. 
Each MLT component of the convolution opens according to \Eq{eq:MLT}.
In the second term on the rhs of \Eq{eq:NMLT}, the set $12$ remains off shell
while there are on-shell propagators in either $1$ or $\overline 2$, and
all the inverted sets from $3$ to $n$ contain on-shell propagators.
For example, at three loops ($n=4$), these convolutions are interpreted as 
\bea
&& {\cal A}^{(2)}_{\rm MLT}(1, 2, 12) \otimes {\cal A}^{(1)}_{\rm MLT}(3, 4) \nn \\
&& \qquad =  \int_{\ell_1,\ell_2, \ell_3} \left( {\cal A}^{(3)}_D(\overline 2, \overline{12}, \overline 4; 1, 3) 
+ {\cal A}^{(3)}_D(1, \overline{12}, \overline 4; 2, 3) \right. \nn \\
&& \qquad \left. + {\cal A}^{(3)}_D(1, 2, \overline 4; 12, 3) + (\overline 4\leftrightarrow 3) \right)~,
\eea
and
\bea
&& {\cal A}^{(1)}_{\rm MLT}(1,2)\otimes {\cal A}^{(0)}(12)\otimes {\cal A}^{(2)}_{\rm MLT}(\overline 3, \overline 4)  \\
&& \qquad = \int_{\ell_1,\ell_2, \ell_3} 
\left( {\cal A}^{(3)}_D(\overline 2, \overline{3}, \overline 4; 1, 12) +
{\cal A}^{(3)}_D(1, \overline{3}, \overline 4; 2, 12) \right)~. \nn 
\eea
The two sets after the semicolon remain off shell.
In total, the number of terms generated by \Eq{eq:NMLT} is $3L-1$.

\begin{figure}[th]
\begin{center}
\includegraphics[width=0.45\textwidth]{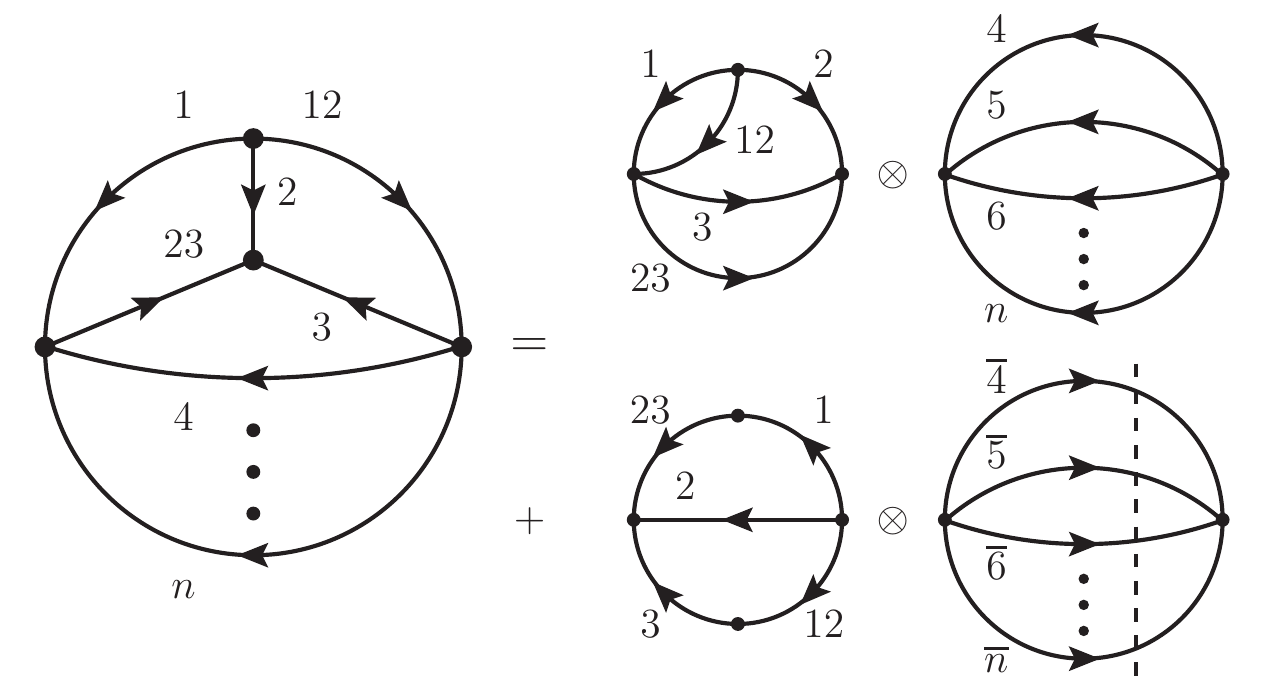}
\caption{Next-to-next-to-maximal loop topology (left) and its convoluted dual representation (right). 
Opening according to \Eq{eq:NNMLT}. 
Only the on-shell cut of the last MLT-like subtopology with reversed momentum flow is shown.
\label{fig:nnmaxtop}}
\end{center}
\end{figure}

Causal thresholds and infrared singularities are then determined by the singular structure 
of the ${\cal A}^{(2)}_{\rm MLT}(1, 2, 12)$ subtopology, and by the singular configurations that split the NMLT 
topology into two disjoint trees with all the on-shell momenta aligned over the causal cut. 
Again, the singular surfaces in the loop three-momenta space are limited by the external momenta, and
all the noncausal singular configurations that arise in individual contributions undergo dual cancellations.

\section{Next-to-next-to-Maximal loop topology}

The last multiloop topology that we consider explicitly is the 
next-to-next-to-maximal loop topology (N$^2$MLT) shown in Fig.~\ref{fig:nnmaxtop}. 
At three loops, it corresponds to the so-called Mercedes-Benz topology. 
Besides the $12$-set, there is another set denoted by $23$ with $q_{i_{23}} = - \ell_2 - \ell_3 + k_{i_{23}}$.
Its LTD representation is given by the following convolution of factorized subtopologies
\bea
&& {\cal A}^{(L)}_{{\rm N}^2{\rm MLT}}(1,\dots,n,12,23) \label{eq:NNMLT} \\ 
&& \qquad = {\cal A}^{(3)}_{\rm NMLT}(1, 2, 3, 12, 23) \otimes {\cal A}^{(L-3)}_{\rm MLT}(4, \ldots, n) \nn \\ 
&& \qquad + {\cal A}^{(2)}_{\rm MLT}(1\cup 23, 2, 3\cup 12) \otimes {\cal A}^{(L-2)}_{\rm MLT}(\overline 4, \ldots, \overline n)~. \nn
\eea
The sets $(1,2,3,12,23)$ form a NMLT subtopology. 
Therefore, the first component of the first term on the rhs of \Eq{eq:NNMLT}  opens iteratively as 
\bea
&& {\cal A}^{(3)}_{\rm NMLT}(1, 2, 3, 12, 23) 
= {\cal A}^{(2)}_{\rm MLT}(1, 2, 12) \otimes {\cal A}^{(1)}_{\rm MLT}(3, 23)  \nn \\
&& + \int_{\ell_1, \ell_2, \ell_3}  \left( {\cal A}^{(3)}_D(1, \overline 3, \overline {23}; 2, 12) + {\cal A}^{(3)}_D(\overline {12}, 3, 23; 1, 2) \right)~. \nn \\
\label{eq:NNMLT2}
\eea
The last two terms on the rhs of \Eq{eq:NNMLT2} are fixed by the condition that the sets $(2, 3, 23)$ cannot generate a disjoint subtree. 
The second term on the rhs of \Eq{eq:NNMLT} contains a two-loop subtopology made of five sets of momenta, 
${\cal A}^{(2)}_{\rm MLT}(1\cup 23, 2, 3\cup 12)$, 
which are grouped into three sets and dualized through \Eq{eq:MLT}. 
For example, propagators in the sets $1$ and $23$ are not set simultaneously on shell.
The number of terms generated by \Eq{eq:NNMLT} is $8(L-1)$. 
As for the NMLT, the causal singularities of the N$^2$MLT topology are determined by its subtopologies 
and by the singular configurations that split the open amplitude into disjoint trees with all the on-shell
momenta aligned over the causal cut. Any other singular configuration 
is entangled among dual amplitudes and cancels.  

We would like to emphasize that \Eq{eq:NNMLT} accounts properly for the NMLT and MLT topologies as well,
if either $23$ or both $12$ and $23$ are taken as empty sets. 
At three loops, therefore, \Eq{eq:NNMLT} emerges as the LTD master topology for opening 
any scattering amplitude from its Feynman representation.

Finally, let us comment on more complex topologies at higher orders. 
For example, let's consider the multiloop topology made of one MLT and two two-loop NMLT subtopologies
that appears for the first time at four loops. This topology opens into nondisjoint trees 
by leaving three loop sets off shell and by introducing on-shell conditions in the others
under certain conditions: either one off-shell set in each subtopology or two in one NMLT subtopology and one in the other
with on-shell propagators in all the sets of the MLT subtopology.  
Once the loop amplitude is open into trees, the singular causal structure is determined by the causal 
singularities of its subtopologies, and all entangled noncausal singularities of the forest cancel.


\section{Conclusions}

We have reformulated the loop-tree duality at higher orders and have obtained very compact open-into-tree 
analytical representations of selected loop topologies to all orders. These loop-tree dual representations exhibit a factorized cascade form 
in terms of simpler subtopologies. Since this factorized structure is imposed by the opening into nondisjoint trees
and by causality, we conjecture that it holds to all loop orders and topologies. 
Remarkably, specific multiloop configurations are described by 
extremely compact dual representations which are, moreover, free of unphysical singularities
and independent of the initial momentum flow. 
This property has been tested with all the topologies and several internal configurations. We also conjecture that 
analytic dual representations in terms of only causal denominators are always plausible. 

The explicit expressions presented in this Letter are sufficient to describe any scattering amplitude up 
to three loops. Other topologies that appear for the first time at four loops and beyond have been anticipated, 
and will be presented in a forthcoming publication.
This reformulation allows for a direct and efficient application to physical scattering processes, 
and is also advantageous to unveil formal aspects of multiloop scattering amplitudes.

{\it Acknowledgements:} 
We thank Stefano Catani for very stimulating discussions. 
This work is supported by the Spanish Government  (Agencia Estatal de Investigaci\'on) and ERDF funds from European
Commission (Grants No. FPA2017-84445-P and No. SEV-2014-0398), Generalitat Valenciana (Grant No. PROMETEO/2017/053),
Consejo Superior de Investigaciones Cient\'{\i}ficas (Grant No. PIE-201750E021) and the COST Action CA16201 PARTICLEFACE. 
JP acknowledges support from "la Caixa" Foundation (No. 100010434, LCF/BQ/IN17/11620037), and the European Union's H2020-MSCA 
Grant Agreement No. 713673; SRU from CONACyT and Universidad Aut\'onoma de Sinaloa; JJAV from Generalitat Valenciana (GRISOLIAP/2018/101);
WJT and AERO from the Spanish Government (FJCI-2017-32128, PRE2018-085925); and
RJHP from Departament de F\'{\i}sica Te\`orica, Universitat de Val\`encia, CONACyT through the project A1-S-33202 (Ciencia B\'asica) and Sistema
Nacional de Investigadores. 


\end{document}